\documentclass[12pt]{article}
\usepackage{a4wide}
\usepackage{amssymb}
\usepackage{graphicx}
\usepackage{amsmath}
\usepackage[center]{subfigure}

\newcommand{\ba}{\begin{eqnarray}}
\newcommand{\ea}{\end{eqnarray}}

\newcommand{\be}{\begin{equation}}
\newcommand{\ee}{\end{equation}}
\newcommand{\GeV}{\rm GeV}
\newcommand{\MeV}{\rm MeV}
\newcommand{\MSbar}{\overline{\rm MS}}
\newcommand{\ep}{\varepsilon}
\newcommand{\ice}[1]{\relax}
\begin{document}
\begin{titlepage}
\begin{flushright}
SI-HEP-2007-19\\
TTP-07-36\\
SFB/CPP-07-83\\
\end{flushright}
\vfill
\begin{center}
{\Large\bf 
Towards NNLO Accuracy in the QCD Sum Rule 
for the Kaon Distribution Amplitude}\\[2cm]
{\large\bf K.~G.~Chetyrkin~$^{(a)}$, 
A.~Khodjamirian~$^{(b)}$, A.~A.~Pivovarov~$^{(b,c,}$}\footnote{partly 
supported by RFFI grant 06-02-16659}$^)$\\[0.5cm]
{\it $^{\mbox{(a)}}$ Institut f\"ur Theoretische Teilchenphysik, Universit\"at
Karlsruhe,\\  D-76128 Karlsruhe, Germany }\\
{\it  $^{\mbox{(b)}}$ Theoretische Physik 1, Fachbereich Physik,
Universit\"at Siegen,\\ D-57068 Siegen, Germany }\\
{\it $^{\mbox{(c)}}$ Institute for Nuclear 
  Research, Russian Academy of Science,\\ 
117312 Moscow, Russia}
\end{center}
\vfill
\begin{abstract}
We calculate the $O(\alpha_s)$ and $O(\alpha_s^2)$ 
gluon radiative corrections to the QCD sum rule for 
the first Gegenbauer moment $a_1^K$ of the kaon 
light-cone distribution amplitude. The NNL0 accuracy is achieved
for the perturbative term 
and quark-condensate contributions to the sum rule.
A complete factorization is implemented, removing
logarithms of $s$-quark mass from the coefficients
in the operator-product expansion.
The sum rule with radiative corrections yields 
$a_1^K(1~\GeV)=0.10\pm 0.04$.  
\end{abstract}
\vfill

\end{titlepage}

\newcommand{\DS}[1]{/\!\!\!#1}

{\bf 1.} Light-cone distribution amplitudes (DA's) of hadrons 
enter various factorization  formulae used for description 
of exclusive processes in QCD. The concept of DA's
allows to describe collinear partons in an 
energetic hadron, separating long-distance dynamics 
from  the  perturbatively calculable hard-scattering 
amplitudes. 

The set of DA's with a growing twist is especially
useful for the pion and kaon because  
their intrinsically small masses make the collinear 
description more  efficient. The lowest twist-2 DA  has a transparent 
physical interpretation, describing  the longitudinal momentum 
distribution in the quark-antiquark Fock-state of a meson. 
Switching from the pion to kaon, one encounters the 
$SU(3)_{fl}$-symmetry violation effects,
which originate from the quark mass difference $m_s-m_{u,d}$. 
These effects have to be accounted 
as  accurate as possible, in order to assess 
the $SU(3)_{fl}$ symmetry relations between the hadronic
amplitudes with pions and kaons. Important examples are 
the relations between $B\to \pi\pi$ and $B\to\pi K, K\bar{K}$ 
charmless decay amplitudes employed in the studies 
of CP-violation and quark-flavour mixing.   

The most essential $SU(3)_{fl}$ -violating 
effects in the kaon twist-2 DA include 
the ratio of the decay constants $f_K/f_\pi$ 
and the difference between the longitudinal momenta of 
strange and nonstrange quark-partons. This
difference is proportional to the  
first moment $a_1^K$ in the decomposition 
of the kaon twist-2 DA in Gegenbauer polynomials, whereas
$a_1^\pi$ vanishes in the isospin ($G$-parity) symmetry limit. 
In addition, the ratio of the second Gegenbauer moments 
$a_2^K/a_2^\pi $ can also deviate from unity;
the effects related to $a^{K}_{n}$
at $n\geq 3$ are usually neglected.

In this paper we concentrate on 
the determination of the asymmetry 
parameter $a_1^K(\mu)$ for the kaon, at a low scale $\mu\sim 1~\GeV$. 
The method originally suggested in~\cite{CZ} and 
based on QCD sum rules~\cite{SVZ} is employed.
The most recent sum rule estimates of $a_1^K$ 
were obtained in~\cite{KMM} and~\cite{BZ}, where, in addition to the 
known leading-order (LO) results,  
the next-to-leading (NLO), $O(\alpha_s)$ correction to 
the 
quark-condensate contribution are taken into account.
These calculations, together with the estimates~\cite{BL,BZ2} 
based on the operator identities, yield the interval
(quoted as a best estimate in~\cite{BBL}):
$
a_1^K(1 \mbox{GeV})=0.06\pm 0.03\, .
$
The positive sign of $a_1^K$ 
corresponds, as expected,  to a 
larger average momentum of the heavier valence $s$-quark 
in the kaon.

The aim of this work is to upgrade the precision of the 
QCD sum rule for $a_1^K$. We calculate 
the gluon radiative corrections to the perturbative 
and quark-condensate contributions 
in NNLO, including $O(\alpha_s)$ and $O(\alpha_s^2)$ 
corrections. This task is technically feasible, 
due to the currently achieved state-of-the-art in 
the calculations of multiloop effects in the two-point 
correlation functions with strange and nonstrange quarks. 
For the correlation function with the scalar 
and pseudoscalar currents the $O(\alpha_s^4)$,
five-loop accuracy has recently been achieved~\cite{BCK}, 
and used, e.g., for the QCD sum rule determination  of the 
strange quark mass~\cite{CK,Jamin,Doming}.
In this case, in the perturbative expansion 
the $O(\alpha_s^2)$  terms are important numerically,
which is one motivation to include these terms
also in the sum rule for $a_1^K$. 
The correlation functions underlying the sum rules 
for Gegenbauer coefficients are however different, because 
the currents contain derivatives. 
Therefore, the calculation 
reported in the present paper, involves
a certain technical novelty. 
In addition, we clarify and take into account the mixing 
of operators that is necessary for the
complete factorization of small and large scales in the 
correlation function. Our result for the first Gegenbauer moment 
of the kaon DA is:
\be
a_1^K(1~\GeV)=0.10\pm 0.04\, .
\label{eq:a1res}
\ee
 
In what follows, we introduce the correlation function, present the expressions
for the new radiative corrections, derive
the resulting QCD sum rule for $a_1^K$, 
including the new $O(\alpha_s)$ and $O(\alpha_s^2)$ terms 
and perform the numerical analysis.\\

{\bf 2.}
The twist-2 DA of the kaon 
enters the standard expression for the 
light-cone expansion of the vacuum-kaon 
bilocal matrix element (we take $K^-$ 
for definiteness): 
\begin{equation}
\langle K^-(q)|\bar{s}(z)\gamma_\mu\gamma_5\left[z,-z\right]u(-z)|0
\rangle_{z^2\to 0}
  = -i q_\mu f_K\int_0^1 du~e^{iu q\cdot z -i\bar{u}q\cdot z}\varphi_K(u,\mu)\,,
\label{eq-phiK}
\end{equation}
where 
the $s$- and $\bar{u}$ quarks carry the momentum fractions 
$u$ and $\bar{u}=1-u$; $[z,-z]$ is the path-ordered gauge-factor
$
[x_1,x_2]=P\exp(i\int_0^1 dv (x_1-x_2)_\rho A^
\rho(vx_1+\bar{v}x_2))\,,
$
and $\mu$ is
the normalization scale determined by the interval $z^2$ near the light-cone.
We use the compact notation $A_\rho =g_s A_{\rho}^ a\lambda^a/2$
for the gluon field and the covariant derivative 
is defined as $D_\rho=\partial_\rho -i A _\rho$.
In (\ref{eq-phiK}), the twist-2 DA $\varphi_K(u)$ is normalized
to unity, so that in the local limit $z\to 0$ one reproduces 
the definition of the kaon decay constant $f_K$.

As usual, $\varphi_K(u)$ is expanded in the 
Gegenbauer polynomials
\begin{equation}
  \varphi_K(u,\mu)=6u\bar{u}\left(1+\sum_{n=1}^\infty a_n^K(\mu)
C_n^{3/2}(u-\bar{u})\right),
\label{eq-moments}
\end{equation}
with the coefficients $a_n^K(\mu)$ (Gegenbauer moments).
The first Gegenbauer moment $a_1^K$ is  proportional to  
the average difference
between the longitudinal momenta of the strange and 
nonstrange quarks in the two-parton state of the kaon. 
Expanding both parts of Eq.~(\ref{eq-phiK}) 
around $z=0$ in local 
operators and using the decomposition~(\ref{eq-moments}) 
with $C_1^{3/2}(x)=3x$
one relates $a_1^K$  to  
the vacuum-to-kaon matrix element 
of a local operator with one derivative: 
\be
\langle K^-(q)|\bar{s}\gamma_\nu\gamma_5i\!
\stackrel{\leftrightarrow}{D}_\lambda\!u|0\rangle
= -iq_\nu q_\lambda f_K \frac35 a_1^K\,,
\label{eq-matrel}
\ee
where $\stackrel{\leftrightarrow}{D}_\lambda
=\stackrel{\rightarrow}{D}_\lambda-\stackrel{\leftarrow}{D}_\lambda$.

The Gegenbauer moments $a_n^{\pi,K}(\mu)$  are known to be 
multiplicatively renormalizable only 
at the one-loop level. Generally, this property 
is lost at higher orders in $\alpha_s$,
e.g., the two-loop renormalization of $a^\pi_2$
calculated in~\cite{Rad_etal} includes operator-mixing effects. 
Still the $a_1^K$ case is special, 
in so far as the underlying operator 
$\bar{s}\gamma_\nu\gamma_5i\!
\stackrel{\leftrightarrow}{D}_\lambda\!u$ 
can only mix with 
$\partial_\lambda (\bar{s}\gamma_\nu\gamma_5 u)$, 
as there is no other local operator
with the same dimension and flavour content.
However, the above two operators have opposite 
$G_{(s)}$-parities, where $G_{(s)}$ is the analog of 
the isospin $G$-parity for the $SU(2)$ subgroups of $SU(3)_{fl}$
involving $s$ quark ($V$ or $U$-spins).
Naturally, $G_{(s)}$-conservation 
is only  realized in the $m_s= m_{u,d}$ limit. Note, however, that 
the ultraviolet renormalization in $\MSbar$-scheme is
a mass-independent procedure. Hence it is legitimate 
to  consider the $SU(3)_{fl}$ limit, while performing 
the renormalization, so that the $G_{(s)}$-conservation protects the 
operators from mixing with each other.
As a result, $a_1^K$ remains multiplicatively renormalizable 
at any order in perturbation theory.
For completeness, we present the well-known expression
for scale-dependence  of $a_1^K$ with the 
two-loop (NLO) accuracy, written in an unexpanded form:
\ba
\label{evol:a1K}
a_1^K(\mu)=
\left(\frac{\alpha_s(\mu)}{\alpha_s(\mu_0)}\right)^\frac{\gamma_0}{\beta_0}
\left(\frac{\beta_0+\beta_1(\alpha_s(\mu_0)/\pi)}{\beta_0+
\beta_1 (\alpha_s(\mu)/\pi)}
\right)^{\left(\frac{\gamma_0}{\beta_0}
-\frac{\gamma_1}{\beta_1}\right)}a_1^K(\mu_0)\,,
\ea
where 
$\gamma_0=8/9$, $\gamma_1=590/243$
are
the anomalous dimensions ~\cite{anomdim:twoloop}  
and $\beta_0=9/4$, $\beta_1=4$ are the 
coefficients of $\beta$-function for $n_f=3$.

As originally suggested in~\cite{CZ}, the few first Gegenbauer 
moments of DA's  can be calculated employing operator-product
expansion (OPE) and QCD sum rules for two-point  vacuum correlation
functions. The method works
well only for the first two coefficients $a_{1,2}^{\pi,K}$. 
In the sum rules for $a_{n\geq3}^{\pi,K}$ the 
condensate contributions grow fast 
with $n$ and the control over OPE is 
lost~\footnote{A
possibility to assess higher Gegenbauer moments
is provided by the model of nonlocal condensates~\cite{RM}.}.  

To obtain a QCD sum rule for $a_1^K$, it is convenient 
to take the 
so called ``diagonal'' correlation function:
\begin{eqnarray}
\Pi_{\mu\nu\lambda}(q)= i\!\int \!d^4x ~e^{i q\cdot x}
\langle 0|T\left\{\bar{u}(x)\gamma_\mu \gamma_5 s(x), 
\bar{s}(0)\gamma_\nu\gamma_5i\stackrel{\leftrightarrow}{D}_\lambda\!u(0)
\right\}|0\rangle= 
q_\mu q_\nu q_\lambda \Pi(q^2)+ \dots,
\label{eq-corr}
\end{eqnarray}
where, for brevity, only the relevant kinematical structure
is shown. In fact, (\ref{eq-corr}) is 
not quite diagonal, because one of 
the operators, the same as in (\ref{eq-matrel}),
contains a derivative. 
A different choice is to correlate the  operator 
in~(\ref{eq-matrel}) with the pseudoscalar current 
$\bar{u}i\gamma_5 s$. This, so called  ``nondiagonal'' 
correlation function was tried in~\cite{BB}, 
but produces an unstable sum rule (see the discussion 
in~\cite{KMM}).\\

\begin{figure}[ht]
\includegraphics[width=5cm]{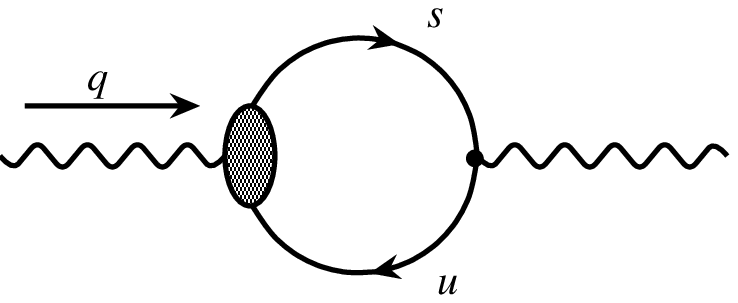}
\includegraphics[width=5cm]{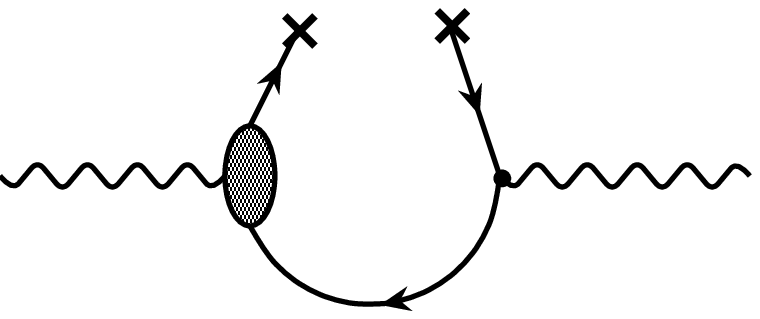}
\includegraphics[width=5cm]{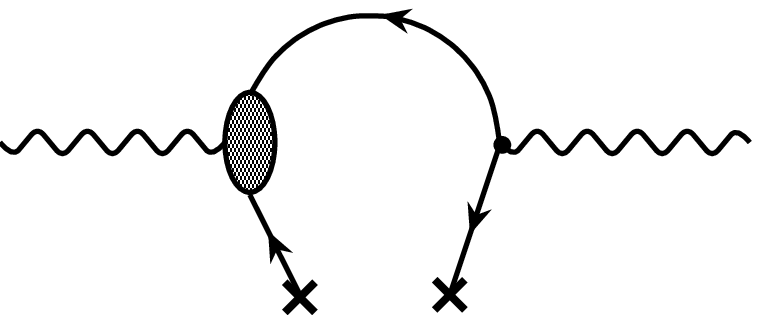}\\
\includegraphics[width=5cm]{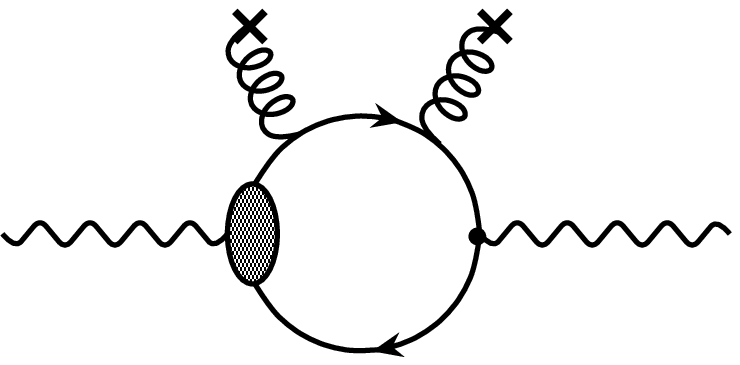}
\vspace{-0.2cm}
\includegraphics[width=5cm]{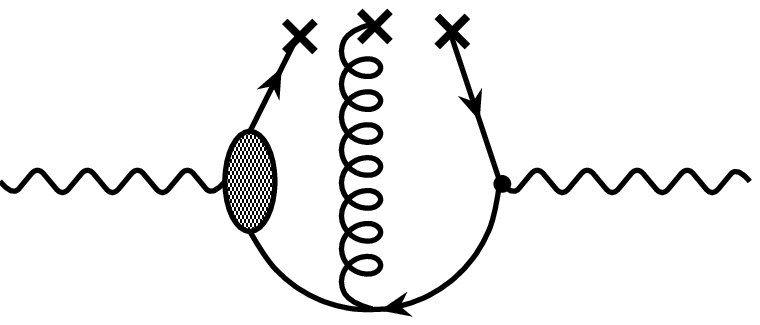}
\includegraphics[width=5cm]{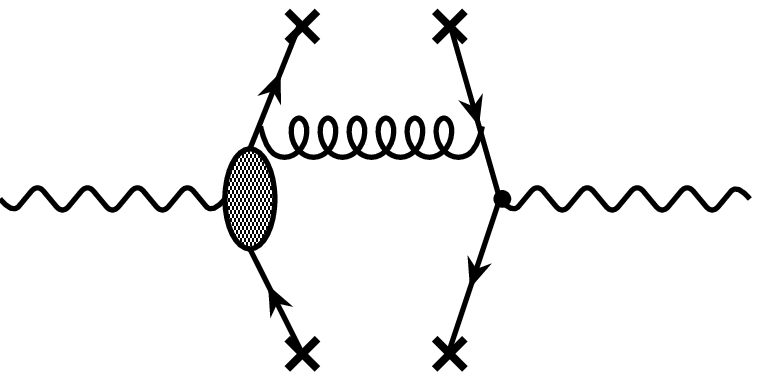}
\vspace{3mm}
\caption{ \it
Diagrams contributing to OPE
of the correlation function~(\ref{eq-corr}) in leading order: 
upper row: perturbative loop and quark-condensate diagrams, 
lower row: ~examples of the gluon-, quark-gluon- and 
four-quark-condensate diagrams. The wavy (curly) lines 
denote external currents 
(gluons), the lines with crosses are vacuum fields. 
The shaded oval distinguishes the current with the derivative.}
\label{fig-diags}
\end{figure}
{\bf 3.} In LO, the operator product expansion (OPE) 
for the correlation function~(\ref{eq-corr}) 
includes the quark-loop diagram at $O(\alpha_s^0)$ 
and the contributions of the vacuum condensates, calculated
in the deep spacelike region,
$Q^2\equiv -q^2 \gg \Lambda_{QCD}^2$.
The corresponding diagrams 
are collected in Fig.~\ref{fig-diags}. 
One obtains a generic expansion
for the  invariant amplitude $\Pi(q^2)$ defined in~(\ref{eq-corr})
in inverse powers of the variable $Q^2$:
\be
\Pi(Q^2,\mu)=\frac{{\cal A}_2(Q^2,\mu)}{Q^2}
+\frac{{\cal A}_4(Q^2,\mu)}{Q^4}
+\frac{{\cal A}_6(Q^2,\mu)}{Q^6}+...\,,
\label{eq-exp}
\ee
where the coefficients ${\cal A}_{d>2}$ contain condensate 
densities with  growing dimensions.
In the above, $\mu$  is the ultraviolet renormalization 
scale in the loop diagrams. The 
OPE is applicable at sufficiently large $Q^2$,
provided that the coefficients ${\cal A}_d$ are proportional
to the powers of the light-quark masses and/or to the 
condensate densities, the latter being of $O(\Lambda_{QCD})$
in some power. 
Hereafter we neglect the $u,d$-quark masses with respect 
to $m_s$. In the expansion (\ref{eq-exp}), the terms proportional 
to $1/Q^{d>6}$ are also neglected, since 
already the contribution of the $d=6$ term is quite small.
Hence, only the vacuum condensates with dimension  $d\leq 6$
are taken into account. 

The gluon radiative corrections can be systematically
included in each term of the OPE~(\ref{eq-exp}). 
Diagrammatically, the $\alpha_s$- ($\alpha_s^2$-) 
corrections correspond to all possible one-gluon (two-gluon) insertions 
in the Fig.~1 diagrams. The second small parameter
entering  the OPE of the correlation function~(\ref{eq-corr})
is the ratio $m_s^2/Q^2$.
Hence, each of the coefficients in~(\ref{eq-exp}) 
can be cast into a form of a generic double expansion:
\ba
{\cal A}_d(Q^2,\mu)=a_{d}^{(0,0)}+
\left(\frac{\alpha_s}{\pi}\right)a_{d}^{(1,0)}+
\left(\frac{\alpha_s}{\pi}\right)^2
a_{d}^{(2,0)}+\left(\frac{m_s^2}{Q^2}\right)a_{d}^{(0,1)}
\nonumber
\\
+\left(\frac{m_s^2}{Q^2}\right)^2a_{d}^{(0,2)}+
\left(\frac{\alpha_s}{\pi}\right)\left(\frac{m_s^2}{Q^2}\right)
a_{d}^{(1,1)}+...\,,
\label{eq-msexp}
\ea
where the coefficients $a_{d}^{(ik)}$ 
multiplying $ (\alpha_s/\pi)^i (m_s^2/Q^2)^k$ 
depend on $\ln (\mu^2/Q^2)\equiv l_Q $.

It is important to assess the numerical role
of  the small parameters in the combined 
expansion~(\ref{eq-msexp}).
We expect to use OPE at $Q^2\simeq 1~\GeV^2$. 
Taking $\alpha_s(1~\GeV)= 0.47$~\cite{PDG}  
and a conservative upper limit for the strange quark mass, 
$m_s(1~\GeV) < 150~\MeV$, one has 
$ m_s^2/Q^2 \leq 0.02 \ll \alpha_s/\pi\simeq 0.15$.
Hence, the perturbative $O(\alpha_s^k)$ contributions to the OPE 
are expected to be more important than the 
$O((m_s^2/Q^2)^k)$ terms with the same power. 
In particular, in the first line of~(\ref{eq-msexp}) 
the second-order, $\alpha_s^2$- correction is expected to 
be of order of the $m_s^2/Q^2$-term. Moreover, 
the observed hierarchy allows one to neglect all ``mixed'' 
$O(\alpha_s^i(m_s^2/Q^2)^k)$ terms with $i,k\neq 0$.
In what follows, we restrict ourselves
to the previously known  first-order corrections 
in $m_s^2/Q^2$ in the $d=2,4$ terms of OPE, 
neglecting them in the $d=6$ term. 

So far only the $O(\alpha_s)$ correction
to the quark-condensate contribution ${\cal A}_4$ 
was calculated~\cite{KMM,BZ}, that is, the coefficient 
$a_4^{(10)}$ in~(\ref{eq-msexp}).
Here we repeat this calculation and, in addition, 
compute the $O(\alpha_s)$ and 
$O(\alpha_s^2)$ terms in  ${\cal A}_2$ and 
the  $O(\alpha_s^2)$ term in ${\cal A}_4$ 
that is, the coefficients $a_2^{(10)},a_2^{(20)}$ and  
 $a_4^{(20)}$, respectively, in~(\ref{eq-msexp}).
Hence, for the largest $d=2,4$ terms of the OPE~(\ref{eq-exp}) 
the NNLO accuracy in $\alpha_s$ is achieved.
For the subleading $d=6$  term in the OPE  
we retain the known LO result. 

The gluon radiative corrections are computed
in the standard $\MSbar$-scheme of renormalization. 
The well developed techniques
of loop calculations are employed, in particular,
the programs QGRAF~\cite{GQRAF}, FORM~\cite{Vermaseren:2000nd}, 
and MINCER~\cite{Gorishnii:1989gt,Larin:1991fz}.
The following 
results are given for $n_f=3$.
Including the new $O(\alpha_s)$ and $O(\alpha_s^2)$ 
contributions, we obtain the $d=2$ term in~(\ref{eq-exp}) originating
from the perturbative contribution:  
\ba
{\cal A}_2(Q^2,\mu)&=&\frac{m_s^2}{4\pi^2}
\Bigg(1+\frac{\alpha_s}{\pi}
\left[\frac{26}{9} + \frac{10}{9}l_Q\right]
\nonumber
\\
&+&\left(\frac{\alpha_s}{\pi}\right)^2
\left[\frac{366659}{11664}-\frac{29}{9}\zeta(3)+\frac{14449}{972}l_Q
+\frac{605}{324}l_Q^2\right]
+3\frac{m_s^2}{Q^2}\left( \frac52+l_Q\right)\Bigg).
\label{eq-a2}
\ea

The $d=4$ contribution in (\ref{eq-exp}) generated by
the quark-condensate term  has the following
expression to $O(\alpha_s^2)$: 
\ba
{\cal A}_4(Q^2,\mu)&=&-m_s \langle \bar{s}s\rangle
\Bigg(1-\frac{\alpha_s}{\pi}\left[\frac{112}{27}+ 
\frac{8}{9}l_Q\right]
\nonumber
\\
&-&\left(\frac{\alpha_s}{\pi}\right)^2
\left[\frac{28135}{1458}-4\zeta(3)
+\frac{218}{27}l_Q
+\frac{49}{81}l_Q^2\right]
+2\frac{m_s^2}{Q^2}\Bigg)
\nonumber
\\
&-&m_s \langle \bar{u}u\rangle\Bigg(
\frac{4\alpha_s}{9\pi}+
\left(\frac{\alpha_s}{\pi}\right)^2\left[\frac{59}{54}+
\frac{49}{81}l_Q\right]\Bigg)\,,
\label{eq-a4}
\ea
where   $\langle \bar{q}q\rangle\equiv \langle 0 |\bar{q}q | 0\rangle$,
($q=s,u$) is the quark-condensate density.

Finally, the $d=6$  term in (\ref{eq-exp})
contains the
LO quark-gluon, gluon and four-quark condensate 
contributions~\cite{KMM,BZ}:  
\be
{\cal A}_6(Q^2,\mu)=\frac{2}{3}{m_s \langle \bar{s}Gs\rangle}+
\frac{1}{3}{m_s^2\langle G^2\rangle}
\left(1+l_Q\right)
-\frac{32}{27}{\pi \alpha_s 
\bigg(\langle \bar{s}s\rangle^2-
\langle \bar{u}u\rangle^2}\bigg)\,,
\label{eq-a6}
\ee
where $\langle \bar s G s \rangle\equiv
\langle 0 |\bar{s}\sigma_{\mu\nu}g_s G^{a\,\mu\nu}(\lambda^a/2) s|0\rangle$,
$\langle G^2\rangle\equiv
\langle  0 |\frac{\alpha_s}{\pi}G^a_{\mu\nu}G^{a\,\mu\nu}|0\rangle$,
and the four-quark condensate is factorized into the products of 
two quark condensate densities, assuming isospin symmetry 
$\langle \bar{d}d\rangle=\langle \bar{u}u\rangle$.

Comparing with the previous calculations~\cite{KMM,BZ},
we observe a difference in the $O(\alpha_s)$  
$s$-quark condensate term. More specifically, the constant term
112/27 in the first line of  (\ref{eq-a4})
has to  replace 124/27 in the corresponding 
expressions obtained in~\cite{KMM} and \cite{BZ}.\footnote{ 
The agreement between \cite{KMM} and our result is restored 
when the renormalization procedure for the one-loop 
diagrams used in~\cite{KMM} is corrected 
by taking into account the $O(\ep)$-term arising 
in $D=4-2\ep$ from the 
tree-level diagram with quark condensate.} 
This difference has, however, only a minor influence on 
the numerical results. 

The form of the coefficient
function multiplying $m_s^4/Q^2$ in (\ref{eq-a2}) and 
$m_s^2\langle G^2\rangle$ in (\ref{eq-a6})
deserves a separate discussion.
One has to emphasize that, in order to achieve  
a complete factorization of small and large  scales in OPE, 
all logarithms of the small parameter $m_s$ 
have to be removed from the coefficient functions
leaving only the powers of $m^2_s/Q^2$.
This procedure, understood long 
ago~\cite{Tkachov:1999nk,BroadGen87,Spiridonov:1988md,Chetyrkin:1994qu,Jamin:1994vr},
generates terms proportional to $\ln(\mu^2/Q^2)$   
instead of $\ln(m_s^2/Q^2)$, e.g.,  in the LO parts of 
${\cal A}_{2}$ and ${\cal A}_{6}$ in 
(\ref{eq-a2}) and (\ref{eq-a6}), respectively. These logarithms  
were not properly treated in previous calculations.
Note that a simple replacement 
$\ln(m_s^2/Q^2)\to \ln (\mu^2/Q^2)$ 
in the coefficient functions can miss a constant term which has to be added 
to the logarithm. Let us, for example, explain 
the calculation of the  
coefficient function in (\ref{eq-a6}), taking into account
the mixing  of ${\langle \bar{s}Gs\rangle}$
and ${m_s\langle G^2\rangle}$ terms under 
renormalization in $\MSbar$ scheme.

We isolate the contributions of the quark-antiquark-gluon 
and gluon condensates to the correlation function
(\ref{eq-corr}) and write them in the following  
convenient form:
\be
\Pi(Q^2)=\frac{2}{3Q^6}\left(C_1(Q^2)m_s\langle \bar{s}Gs\rangle
+\frac{1}{2}C_2(Q^2){m_s^2\langle G^2\rangle}\right)+\ldots
\label{eq-piform}
\ee
where all other contributions indicated  by ellipses
are not important for this discussion. 
The two terms shown in (\ref{eq-piform}) 
are generated by the renormalized 
$d=6$ operators $m_s\bar s G s$ and $m_s^2G^2$. 
To calculate the coefficient 
functions $C_{1,2}(Q^2)$  at $D=4-2\ep$  
it is sufficient to write the pattern of the mixing of these two 
operators in the form 
\be 
m_s\bar{s}Gs=\Bigg[m_s\bar{s}Gs+\frac{1}{2\ep}m^2_s G^2\Bigg]_{nr}\,,
\label{eq-bare}
\ee
where the index  $"nr"$ indicates that the operators 
on r.h.s. are constructed from bare, non-renormalized
quark and gluon fields. The $Z$ factor of the multiplicative
renormalization of $(\bar{s}Gs)_{nr}$  can be put to unity
in this approximation. To proceed, we need the expansion 
of the operator product in~(\ref{eq-corr}) before 
the vacuum average is taken:
\ba
i\int d^4x ~e^{i q\cdot x}
T\{\bar{u}(x)\gamma_\mu \gamma_5 s(x), 
\bar{s}(0)\gamma_\nu\gamma_5i\stackrel{\leftrightarrow}{D}_\lambda\!u(0)\}
\nonumber\\
= \frac{2}{3Q^6}\left(C_1(Q^2)
\Big[m_s\bar{s}Gs+\frac{1}{2\ep}m_s^2 G^2\Big]_{nr}
+\frac{1}{2}C_2(Q^2) \Big[m_s^2 G^2\Big]_{nr}\right)+\ldots
\label{pre-OPE}
\ea
where we use (\ref{eq-bare}).
The coefficient functions $C_{1,2}(Q)$ can now be 
obtained by projecting both parts of 
(\ref{pre-OPE}) onto the suitable 
free quark-gluon states.
Sandwiching (\ref{pre-OPE}) between 
the $\langle\bar{s}g |$  and $|s\rangle$ states, 
and computing the relevant diagrams we find:
\be 
\label{eq:C1}
C_1=1+\frac{7}{6}\ep \,.
\ee
Note that one has to retain the $O(\ep)$ term for the finite 
tree-level diagrams. Furthermore, projecting (\ref{pre-OPE}) 
onto the two-gluon state, we obtain:
\be
\label{eq:C2}
\frac{1}{\ep}C_1+C_2=\frac{1}{\ep}+\frac{13}{6}+l_Q\,.
\ee
After substituting $C_{1}$ from (\ref{eq:C1}) to
(\ref{eq:C2}), the  $1/\ep$ poles cancel each other 
and the desired results for the renormalized 
coefficient functions are obtained at $\ep\to 0$: 
\be 
\label{eq:C1C2}
C_{1}(Q^2)=1,~~C_2(Q^2)=1+l_Q\,.
\ee
We again emphasize that one has to keep the 
$O(\ep)$ term in the expression for the 
tree-level coefficient function $C_{1}(Q^2)$ in order to get the 
non-logarithmic contribution to 
$C_2(Q^2)$. 
The absence of $1/\ep$ poles in the above calculation 
can also be interpreted as a result of the 
cancellation between ultraviolet 
and infrared divergences. Indeed, the $1/\ep$ pole
in the mixing of operators emerges as an ultraviolet
divergence, whereas $1/\ep$ in the loop diagrams used to
calculate the coefficient functions has an 
infrared origin.
A similar derivation is applied to the mixing 
of $O(m_s\langle \bar s s\rangle)$
and $O(m_s^4)$ terms in OPE, yielding the constant 
term $5/2$ that accompanies
$\ln(\mu^2/Q^2)$ in the coefficient function (\ref{eq-a2}).\\

{\bf 4.}
After the correlation function is calculated,
the derivation of the sum rule follows the standard
procedure. The OPE (\ref{eq-exp}) with the coefficients 
given in (\ref{eq-a2}),(\ref{eq-a4})
and (\ref{eq-a6}) is equated to the hadronic dispersion relation
\begin{equation}
\Pi(q^2)= 
\frac {\frac35 a_1^K f_K^2}{m_K^2-q^2}
+
\int \limits_{s_h}^\infty ds \frac{\rho^h(s)}{s-q^2}\,.
\label{eq-disp}
\end{equation}
In the above, possible subtractions are ignored 
in anticipation of the Borel transformation.  
The residue of the kaon pole is obtained  
by combining  the matrix element (\ref{eq-matrel}) with the  
definition of the kaon decay constant.

The spectral density $\rho^h(s)$ includes  the  
contributions of hadronic continuum and resonances  
with $J^P=0^-,1^+$ and strangeness:
$K\pi\pi$, $ K^*\pi$, $K\rho$, $K_1(1270)$, $K_1(1400)$,... .
Accordingly, the lower limit of integration 
is $s_h=(m_K+2m_\pi)^2$, the invariant 
mass squared of the lightest continuum state in this channel. 
To approximate  $\rho^h(s)$, we employ 
the quark-hadron duality approximation: 
\be
\rho^h(s)\Theta(s-s_0^h)=
\rho^{OPE}(s)\Theta(s-s_0^K)\,,
\label{eq-spectr}
\ee
where $s_0^K$ is the effective threshold, 
and the spectral density:
\ba
\rho^{OPE}(s,\mu)=\frac{1}{\pi}{\rm Im}\Pi(s,\mu)=
-\frac{m_s^2}{4\pi^2s}\Bigg(
\frac{10\alpha_s}{9\pi}
+\left(\frac{\alpha_s}{\pi}\ \right)^2\Big[\frac{14449}{972}+
\frac{605}{162}l_s\Big]-\frac{3m_s^2}{s}\Bigg)
\nonumber\\
+\frac{m_s\langle \bar{s}s\rangle}{s^2}
\Bigg(\frac{8\alpha_s}{9\pi}+
\left(\frac{\alpha_s}{\pi}\ \right)^2\Big[
\frac{218}{27}+\frac{98}{81}l_s\Big]\Bigg)
-\frac{m_s\langle \bar{u}u\rangle}{s^2}
\frac{49}{81}\left(\frac{\alpha_s}{\pi}\right)^2
-\frac{m_s^2\langle G^2 \rangle}{3s^3} 
\label{eq-rho}
\ea
with $l_s\equiv \ln(\mu^2/s)$, 
is obtained by calculating  the imaginary part 
of $\Pi(Q^2)$ at positive $s=-Q^2$. In the above, 
the running parameters ($\alpha_s$, $m_s$) 
are taken in the  $\MSbar$ scheme and 
normalized at the  scale $\mu$. 

The next step is the Borel transformation 
of (\ref{eq-disp}), which
eliminates the subtraction terms and suppresses 
the integral over $\rho^h(s)$ , so that
the resulting relation becomes 
less sensitive to the duality approximation.
The transformed invariant amplitude $\Pi(q^2)$ 
has the following form:  
\ba
\Pi(M^2)&=&\frac{m_s^2}{4\pi^2}
\Bigg(1+\frac{\alpha_s}{\pi}
\left[\frac{26}{9} + \frac{10}{9}(l_M+\gamma_E)\right]
+\left(\frac{\alpha_s}{\pi}\right)^2
\bigg[\frac{366659}{11664}+\frac{14449}{972}(l_M+\gamma_E)
\nonumber
\\
&+&\frac{605}{324}\Big((l_M+\gamma_E)^2-\frac{\pi^2}{6}\Big)-
\frac{29}{9}\zeta(3)\bigg]
+3\frac{m_s^2}{M^2}\left[\frac32+ l_M+\gamma_E\right]\Bigg)
\nonumber
\\
&-&\frac{m_s\langle\bar{s}s\rangle}{M^2}
\Bigg(1-\frac{\alpha_s}{\pi}\left[\frac{88}{27}+ 
\frac{8}{9}(l_M+\gamma_E)\right]
-\left(\frac{\alpha_s}{\pi}\right)^2
\bigg[\frac{18127}{1458}+\frac{556}{81}(l_M+\gamma_E)
\nonumber
\\
&+&\frac{49}{81}\bigg((l_M+\gamma_E)^2
-\frac{\pi^2}{6}\bigg)-4\zeta(3)\bigg]+\frac{m_s^2}{M^2}\Bigg)
\nonumber
\\
&-&\frac{m_s \langle \bar{u}u\rangle}{M^2}\Bigg(
\frac{4\alpha_s}{9\pi}+
\left(\frac{\alpha_s}{\pi}\right)^2\left[\frac{79}{162}
+\frac{49}{81}(l_M+\gamma_E)\right]\Bigg)
\nonumber
\\
&+&\frac{m_s \langle \bar{s}Gs\rangle}{3M^4}+
\frac{m_s^2\langle G^2\rangle}{6M^4}
\left(-\frac{1}{2}+l_M+\gamma_E\right)
-\frac{16\pi\alpha_s}{27 M^4} 
\bigg(\langle \bar{s}s\rangle^2-
\langle \bar{u}u\rangle^2\bigg)\,,
\label{eq-Borel}
\ea
where $l_M\equiv \ln(\mu^2/M^2)$ and 
$\gamma_E$ is the Euler constant.
Finally, the sum rule for the kaon Gegenbauer moment $a_1^K$
obtained from  (\ref{eq-disp}) using (\ref{eq-spectr}) reads:
\be
a_1^K = \frac{5}{3f_K^2}e^{m_K^2/M^2}\Bigg(\Pi(M^2)-
\int \limits_{s_0^K}^\infty ds \rho^{OPE}(s)e^{-s/M^2}
\Bigg)\,,
\label{eq-SR}
\ee
where the functions $\Pi(M^2)$ and $\rho^{OPE}(s)$ 
are given in (\ref{eq-Borel}) and (\ref{eq-rho}), respectively.
An equivalent form of the sum rule where the entire r.h.s. 
is represented as a duality integral over the spectral density,
was used in~\cite{KMM,BZ}.\\

{\bf 5.}
To perform the numerical analysis of the sum rule~(\ref{eq-SR}), 
we specify the relevant input parameters, starting from the kaon 
mass $m_K^{\pm}=493.58~\MeV$ and decay constant 
$f_K=159.8\pm 1.4\pm 0.44~\MeV $ \cite{PDG}.
For the strange quark mass we adopt: 
\be
m_s(\mbox{2 GeV})= 98\pm 16~\mbox{MeV}  
\label{eq:smass}
\ee
which covers the intervals 
$m_s(\mbox{2 GeV} )= 105\pm 6 \pm 7~\mbox{MeV}$~\cite{CK}
(with uncertainties added in quadrature) 
and $ m_s(\mbox{2 GeV} )=92\pm 9~\mbox{MeV}$~\cite{Jamin}
from the recent 
QCD sum rule determinations with $O(\alpha_s^4)$ accuracy.

The running of the coupling 
$\alpha_s(m_Z)=0.1176\pm 0.002$~\cite{PDG} to the 
lower scale $\mu$  
is taken with the 4-loop accuracy applying the program 
presented in~\cite{RunDec}. 
The default value of the renormalization scale is 
$\mu=1~\GeV$, where the expansion parameter is 
$\alpha_s(1~\mbox{GeV})/\pi=0.15\pm 0.01$ and  
$m_s(\mbox{1 GeV})= 128\pm 21~\mbox{MeV}$ is obtained 
from~(\ref{eq:smass}). 

The nonstrange quark condensate density 
$\langle \bar q q \rangle$ 
($q=u,d$) is fixed from the Gell-Mann-Oakes-Renner relation
$\langle \bar q q \rangle = -m_\pi^2f_\pi^2/[2(m_u+m_d)]$.
Taking the non-lattice averages for the $u$- and $d$-quark masses,
$m_u(2~\GeV)=3\pm 1~\MeV$ and $m_d(2~\GeV)=6.0\pm 1.5~\MeV$, 
together with $f_\pi=130.7\pm 0.1\pm 0.36~\MeV$ and 
$m_\pi=139.57~\MeV$ from~\cite{PDG},
we get $\langle \bar q q (2~\GeV)\rangle
=-(0.264^{+0.031}_{-0.020}~\GeV)^3$,
and correspondingly,  $\langle \bar q q (1~\GeV)\rangle
=-(0.242^{+0.028}_{-0.019}~\GeV)^3$.
As mentioned above, in all other formulae 
the light-quark masses are neglected. Furthermore, the ratio of strange and 
nonstrange condensates $\langle \bar s s \rangle/\langle \bar q q \rangle 
= 0.8\pm 0.3$ is adopted~\cite{Belyaev:1982cd,ssqq,Ioffe}.

The accuracy of the condensate densities with higher dimensions 
is less  important. In particular, we take for the quark-gluon condensate
density the standard parameterization
$\langle \bar{s}Gs\rangle 
=m_0^2 \langle \bar{s}s\rangle(1~\mbox{GeV})$  
with $m_0^2=0.8\pm 0.2~\GeV^2$~\cite{Ovchinnikov:1988gk,Ioffe}
(neglecting the running), and a very 
wide interval for the gluon condensate density 
$\langle G^2\rangle=0.012\pm 0.012~\GeV^4$. We also allow for a 
varying factor $0.1\div 1$ multiplying the square of the quark condensate 
density in the factorization relation for the four-quark 
condensates.

The Borel parameter interval $M^2 = 1.0-2.0~\mbox{\GeV}^2$ is adopted, 
so that 
the $d=6$ contribution in $\Pi(M^2)$ 
remains lower than 30\%. The upper limit is taken to avoid too 
large contributions of the excited and continuum  states 
estimated with duality approximation. 
The dependence of $a_1^K$ on $M^2$ for $s_0^K=1.05~\GeV^2$ 
and at central values
of all input parameters is plotted in Fig.~2, showing a 
remarkable stability. There we also plot separate contributions
to the sum rule originating from the $d=2,4,6$ terms of the expansion
(\ref{eq-exp}).
The threshold parameter $s_0^K=1.05~\GeV^2$
was fixed in~\cite{KMM} from the sum rule for $f_K$.
In fact, the $s_0^K$-dependence of the sum rule result 
turns out to be rather weak,
if one varies the threshold in rather wide limits 
$s_0^K=0.9 - 1.4~\GeV^2$, the dependence of  
$a_1^K$ on $M^2$ remaining flat. 
The reason is that the spectral density  (\ref{eq-rho}) 
is small in the region above the duality interval for the kaon. 
Therefore, although the pattern of hadronic states 
(resonances and continuum states) 
in this region  is rather complicated, we 
expect that the integral over these states is also small. Hence 
the sum rule result will not  noticeably change  
if one modifies the duality ansatz for the hadronic spectral 
density, e.g. by adding more resonances
and correspondingly increasing the threshold parameter
(as shown e.g., in  \cite{KMM}).   

\begin{figure}[t]
\label{fig:a1K}
\begin{center}
\includegraphics[width=.5\textwidth]{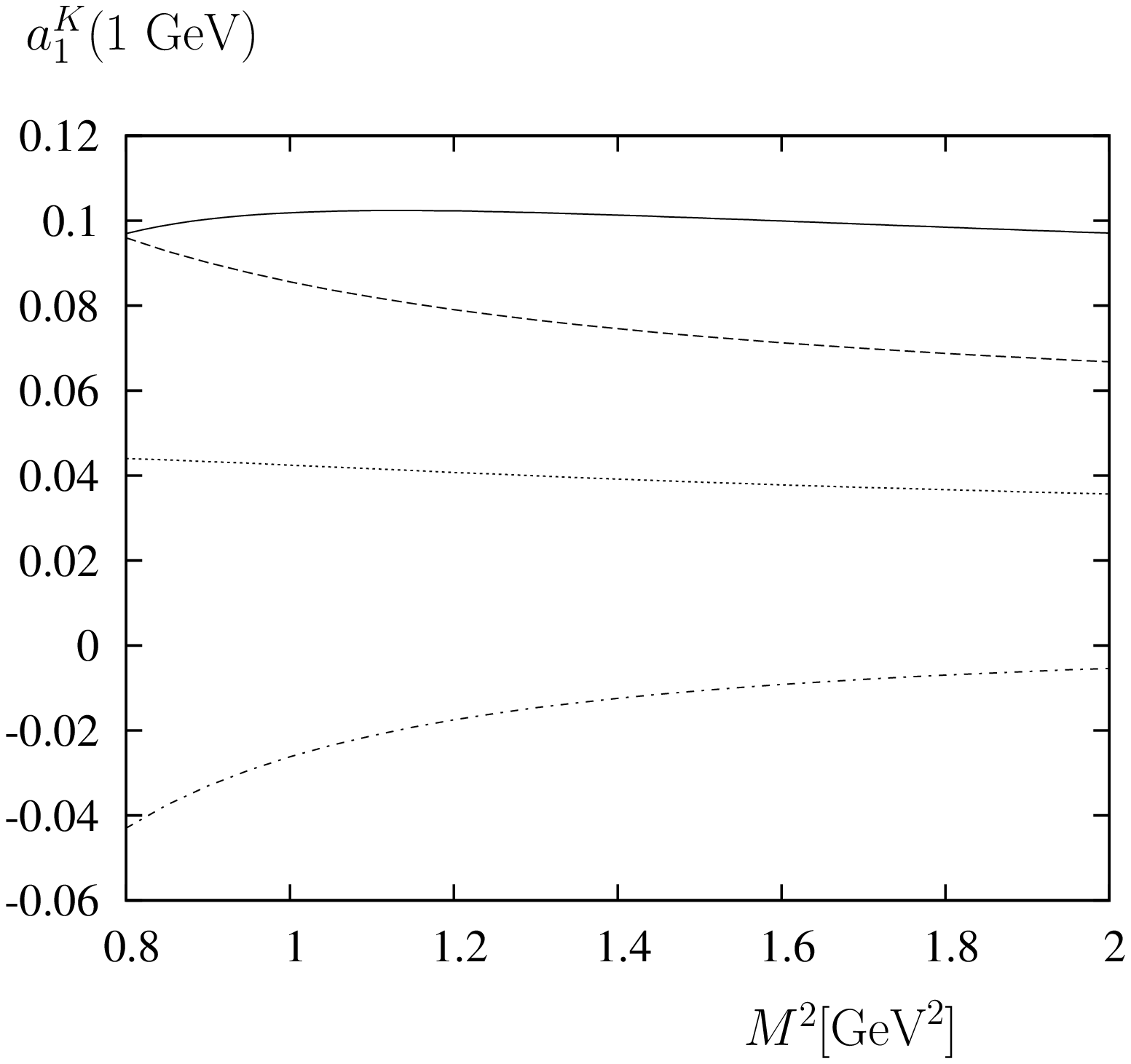}
\end{center}
\vspace{-0.5cm}
\caption{\it  The kaon Gegenbauer moment
$a_1^K(1~\mbox{GeV})$ calculated from the QCD 
sum rule as a function  of the Borel parameter (solid);
the contributions of $d=2$, $d=4$ and $d=6$ terms to (\ref{eq-exp}) are 
shown with  dashed, dotted and dash-dotted lines, respectively.}
\end{figure}

The numerical prediction of the sum rule 
is represented in the form: 
\ba
\label{final:a1Kfinal}
a_1^K(1~\GeV)=0.100\pm 0.003|_{\rm SR}
\pm 0.003|_{\alpha_s}\pm 0.035|_{m_s}
\pm 0.022|_{m_q}\pm 0.013|_{cond}\,,
\ea
where the first (``SR'') error is the combined uncertainty of $a_1^K$ 
due to
the variation of  $M^2$ and $s_0^K$, representing a sort of an 
``intrinsic '' 
uncertainty of the sum rule. 
The subsequent errors correspond to the individual variations 
of $\alpha_s(m_Z)$, $m_s(2~\mbox{GeV})$ and 
$m_{u,d}(2~\mbox{GeV})$ (that is, $\langle \bar{q}q\rangle$),
within the adopted intervals. The last error (``cond'') 
shown in (\ref{final:a1Kfinal}) is a combined uncertainty
due to variation of all remaining condensate parameters.
Finally, adding the individual uncertainties in quadrature 
we obtain the interval~(\ref{eq:a1res}).

Let us now discuss the structure of the perturbative series 
as it follows from the numerical analysis. First, we observe 
that gluon radiative  corrections substantially enhance the
perturbative $d=2$, $O(m_s^2)$ term and suppress 
the quark condensate term. Numerically,  
for the $O(m_s^2)$ term
in (\ref{eq-Borel})
at $\mu=M=1~\GeV$ the following $\alpha_s$-expansion 
is obtained (in ${\rm \overline{MS}}$-scheme):  
\be
\Pi^{(m_s^2)}(1~{\GeV})=\frac{ {m}_s^2(1~\GeV)}{4\pi^2}
\left[1 + 3.53 \left(\frac{\alpha_s(1~\GeV)}{\pi}\right) + 
33.7\left(\frac{\alpha_s(1~\GeV)}{\pi}\right)^2\right]\,,
\label{eq:pims2}
\ee
revealing a poor convergence in  $\alpha_s$
at $\alpha_s(1~\GeV)/\pi\simeq 0.15$.
It is not surprising, if one recalls 
the higher-order perturbative corrections 
calculated for the $m_s$ determination  
from $\tau$ decays or
from QCD sum rules based on scalar/pseudoscalar correlation 
functions, where the situation is in fact similar. 
For instance,  in the $O(m_s^2)$ part of the Borel-transformed 
scalar/pseudoscalar correlation function \cite{BCK,CK} 
one has (see e.g., eq.(16) in \cite{CK}):
\ba
\Pi^{(5)''(m_s^2)}(1~{\GeV})=
&&
\frac{3 {m}_s^2(1~\GeV)}{8\pi^2}
\Bigg[1 + 4.821 \left(\frac{\alpha_s(1~\GeV)}{\pi}\right) 
+ 21.98\left(\frac{\alpha_s(1~\GeV)}{\pi}\right)^2
\nonumber\\
+&& 
53.1\left(\frac{\alpha_s(1~\GeV)}{\pi}\right)^3
+ 31.6\left(\frac{\alpha_s(1~\GeV)}{\pi}\right)^4\Bigg]
\nonumber\\
\simeq  
&&
\frac{3 {m}_s^2(1~\GeV)}{8\pi^2}\left(1 + 0.72 + 0.49+0.18 
+ 0.02 \right).
\label{eq:pimsms2}
\ea
The ratio of  (\ref{eq:pimsms2}) (taken with $O(\alpha_s^2)$ 
accuracy) and (\ref{eq:pims2}) has a much better convergence:
\be 
\frac{\Pi^{(5)''(m_s^2)}(1~{\GeV})}{\Pi^{(m_s^2)}(1~{\GeV})}
=\frac{3}{2}
\left[1 +  1.29 \left(\frac{\alpha_s(1~\GeV)}{\pi}\right) 
- 16.2\left(\frac{\alpha_s(1~\GeV)}{\pi}\right)^2+\ldots\right]\,,
\ee
where the corrections cancel each other to a  great extent.

In the quark condensate contribution, 
the radiative corrections are less sizable:
at $\mu=1~\GeV$ the numerical 
hierarchy of the terms multiplying the strange quark 
condensate density in (\ref{eq-a4}) reads:  
\be
\Pi^{(m_s\langle \bar{s}s\rangle)}(1~{\GeV})=
m_s\langle \bar s s \rangle \left(1 
- 3.77\left(\frac{\alpha_s(1~\GeV)}{\pi}\right) 
-10.8\left(\frac{\alpha_s(1~\GeV)}{\pi}\right)^2 \right)\,,
\ee
where we again find that the second-order correction in 
$\alpha_s$ is important. 

We conclude that after taking into account the NNLO 
perturbative corrections, the numerical 
pattern of the sum rule for $a_1^K$ drastically changes: the 
coefficient of the $d=2$ term gets enhanced, whereas the 
$d=4$ term decreases. Note also that including  
higher-order perturbative corrections in the sum rule, 
makes more consistent the use of the input parameters, such 
as $m_s$, determined with a high accuracy, 
up to $O(\alpha_s^4)$.

Finally, the $d=6$ subleading contributions to the sum rule
play an important role in providing the  Borel stability.
The mixed quark-antiquark-gluon condensate 
dominates numerically in~(\ref{eq-a6}) 
yielding a negative contribution to the
Borel-transformed correlation function and stabilizing 
the whole sum of OPE terms.
Computation of the radiative corrections to this term
is a difficult task beyond our scope. 
Having in mind the large uncertainty of the mixed condensate
density, we expect that the $\alpha_s$-corrections to 
the coefficient function of the $\langle \bar{s}Gs\rangle$ term 
will hardly improve the overall accuracy of its contribution.  

Finally, we checked that due to the enhanced precision 
in $\alpha_s$, the dependence of the sum rule prediction
for $a^K_1$ on the renormalization scale $\mu$ becomes small. 
Taking $\mu>1$ GeV we calculate 
$a_1^K(1~\GeV)$ by rescaling the sum rule result  
with the NLO scale dependence (\ref{evol:a1K}). 
The LO and NLO logarithmic dependences naturally cancel out 
in $d=2,4$ terms, leaving a very mild residual scale-dependence 
due to the unaccounted $\alpha_s$-correction to the 
$\langle \bar{s}Gs\rangle$ term.\\

{\bf 6.}
Concluding, we have calculated the NNLO gluon radiative corrections 
to the QCD sum rule for the first Gegenbauer moment of the kaon
distribution amplitude. The corrections turned out
to be numerically important, they change the relative
magnitude of the $d=2$ (loop diagrams) and $d=4,6$ 
(condensate) terms in the OPE,
improving also the Borel stability of the sum rule. 

The uncertainty of $a_1^K$ is still large and amounts up to 40\%,
due mainly to the limited  precision
of the light quark masses: $m_s$ directly entering 
the sum rule and  $m_{u,d}$ determining 
the quark-condensate densities via 
Gell-Mann-Oakes-Renner relation. 
A better determination of the ratio of strange and 
nonstrange condensates and of the mixed quark-gluon condensate 
are another possibilities of reducing the theory error. 
The calculation of the radiative corrections to 
the quark-gluon condensate contribution can also improve 
the accuracy. This however requires
a dedicated computational effort including an analysis of the 
whole basis of the dimension-six operators.
The weak dependence of the sum rule on the threshold parameter 
$s_0^K$ indicates that the one-resonance (kaon) duality ansatz is 
quite satisfactory. Still, an additional analysis,   
including the axial-vector and radially excited kaon resonances
could  provide a better understanding of the duality pattern in this
channel.

Our result for $a_1^K$ is somewhat larger than  
the previous estimates ~\cite{KMM,BZ,BBL}, and the uncertainty, 
we believe, is more realistic.
For comparison we also quote the two recent lattice QCD determinations
of this parameter: 
$a_1^K(2~\GeV)= 0.0453\pm 0.0009\pm 0.0029$~\cite{latt1}  
and $a_1^K(2~\GeV)= 0.048\pm 0.003$~\cite{latt2}, which have 
achieved  a rather small error.
By evolving our result (\ref{eq:a1res}) to this scale 
with the help of the scale-dependence
(\ref{evol:a1K}) we find $a_1^K(2~\GeV)=0.08\pm 0.04$, that is, 
within uncertainties, only a marginal agreement with the lattice
results.
Let us note that both lattice determinations
use a linear extrapolation in $m_s$ (kaon mass squared) 
inspired by ChPT  in the leading order 
(see also \cite{ChenStewart}). We would like to stress again 
that in our analysis the contribution to the sum rule proportional 
to $m_s^2$ is  enhanced while 
the term proportional to $m_s\langle \bar s s\rangle$ is suppressed,
both enhancement and suppression being caused by the radiative 
corrections. In the language of ChPT, this observation
could indicate an important role of the next-to-leading
terms in  the expansion of $a_1^K$ in the kaon mass.\\

{\bf Acknowledgements}

We are grateful to V.~Braun, Th.~Mannel, and R.~Zwicky for useful discussions,
and to M.~Melcher for collaboration at the initial stage 
of this project. 
The work of K.G.Ch. was supported by the DFG 
Sonderforschungsbereich  SFB/TR-9 
(``Computational Particle Physics''), the work of A.K. 
was supported  by the DFG project KH 205/1-2. 
A.A.P. acknowledges the hospitality of  the Particle 
Theory Group at Siegen University 
where this work was done during his stay as a 
DFG Mercator Guest Professor (Contract DFG SI 349/10-1).
The work of A.A.P. was supported in part by the RFFI
grant 06-02-16659.

\end{document}